\documentclass[twocolumn,superscriptaddress]{revtex4-2}
\usepackage[colorlinks=true, citecolor=blue, urlcolor=blue, linkcolor=red]{hyperref}
\usepackage{amsmath,amssymb,tikz,scalerel,physics}
\hypersetup{colorlinks=true, linkcolor=red, citecolor=blue, urlcolor=blue}
\usetikzlibrary{svg.path}
\definecolor{orcidlogocol}{HTML}{A6CE39}
\tikzset{
	orcidlogo/.pic={
		\fill[orcidlogocol] svg{M256,128c0,70.7-57.3,128-128,128C57.3,256,0,198.7,0,128C0,57.3,57.3,0,128,0C198.7,0,256,57.3,256,128z};
		\fill[white] svg{M86.3,186.2H70.9V79.1h15.4v48.4V186.2z}
		svg{M108.9,79.1h41.6c39.6,0,57,28.3,57,53.6c0,27.5-21.5,53.6-56.8,53.6h-41.8V79.1z M124.3,172.4h24.5c34.9,0,42.9-26.5,42.9-39.7c0-21.5-13.7-39.7-43.7-39.7h-23.7V172.4z}
		svg{M88.7,56.8c0,5.5-4.5,10.1-10.1,10.1c-5.6,0-10.1-4.6-10.1-10.1c0-5.6,4.5-10.1,10.1-10.1C84.2,46.7,88.7,51.3,88.7,56.8z};}}
\newcommand\orcid[1]{\href{https://orcid.org/#1}{\mbox{\scalerel*{\begin{tikzpicture}[yscale=-1,transform shape]\pic{orcidlogo};\end{tikzpicture}}{|}}}}

\begin{document}
\title{FR I Radio Galaxy 3C 348 Accumulation Patterns and Dust Ring Occlusion Effects}

\author{Yu-Jia Zhao}
\author{Jing-fu Hu}
\email{3104034484@qq.com}
\affiliation{School of Physics and Electronic Information Engineering, Qinghai Normal University, Xining, 810000, China}

%\affiliation{Key Laboratory of Quantum Theory and Applications of MoE \& Key Laboratory of Theoretical Physics of Gansu Province, Lanzhou University, Lanzhou 730000, China}
%\author{Jun-Hong An\orcid{0000-0002-3475-0729}}
%\email{anjhong@lzu.edu.cn}
%\affiliation{School of Physical Science and Technology \& Lanzhou Center for Theoretical Physics, Lanzhou University, Lanzhou 730000, China}
%\affiliation{Key Laboratory of Quantum Theory and Applications of MoE \& Key Laboratory of Theoretical Physics of Gansu Province, Lanzhou University, Lanzhou 730000, China}

\begin{abstract}
Radio galaxies can be classified into two types, FR I and FR II, depending on their morphology. So far, the reasons for the different behaviour of FR I and FR II in observations have not been clarified. While the Unified Model suggests that the viewing angle and the obscuring effect of the dust ring are the main reasons for the difference in the classification of FR I and FR II radio galaxies, it has been found that the accretion rate and the accretion pattern of the central engine may play a more crucial role. In order to investigate the physical properties of these sources, the cloudy program was used to simulate the 3C 348 radio source outlook ionisation model, and it was found that its optical emission in the nuclear region is significantly lower than the theoretical prediction, which suggests the existence of an obscuration effect, which was later further proved by the spectral decomposition method. And using the thermometric photometry-jet relationship, it is found that the accretion rate in its nuclear region is much lower than the accretion rate inverted from the total jet power, indicating that its accretion mode is in the critical stage of the transition from the standard thin disc to the ADAF. Combined with its Bondi accretion rate, the fitted energy spectrum of the nuclear region shows that the evolution of its jet power is controlled by the change of the central accretion engine.

\noindent \textbf{Keywords:} active galactic nuclei, radio galaxies, accretion mode, trous, jet power
\end{abstract}
\maketitle

\section{Introduction}

Based on the spectral emission characteristics of galaxies and the physical properties of their nuclear regions, active galactic nuclei can be classified into different subtypes such as quasars, Seyfert galaxies, flare variables, low ionisation nuclear emission regions (LINERs), radio galaxies, and so on. In the so-called standard unifying framework of active galactic nuclei, these differences in classification are not due to differences in their intrinsic physical properties, but rather to differences in accretion rates and in the angle between the line of sight and their jets~\cite{urry1995unified}. Specifically, high-luminosity quasars and low-luminosity low-ionisation nuclear emitting regions correspond to two typical accretion states, respectively: the former in the form of a geometrically thin, optically thick standard accretion disc, and the latter in the geometrically thick, optically thin, radially shifted dominated accretion flow (ADAF) mode~\cite{kato1998black,narayan1994advection,narayan1995advection(a),yuan2014hot}. In addition, flares, Seyfert galaxies, and radio galaxies can be observed by the angle between the jet direction and the line-of-sight direction, which is very large for radio galaxies and Seyfert galaxies and very small or even in the same plane for flares. It is thus clear that the dust ring is a key fundamental structure in the theoretical architecture of the unified model of active galactic nuclei. Moreover, the theory suggests that the different properties of active galactic nuclei due to differences in the observational line of sight are rooted in the obscuring effect of the dust rings.

The results of the 1996 Ledlow \& Owen study of the collected sample show that the correlation between the radio luminosities and optical magnitudes of FR I radio galaxies and FR II radio galaxies is not significant, yet there is an extremely important functional relationship between the FR I and FR II excesses. We are able to clearly distinguish between these two types of radio galaxies with the help of a linear function in the plane of the optical magnitude of the host galaxy and the total radio magnitude.~\cite{ledlow199620cm} The majority of FR II radio galaxies are distributed above this line of demarcation, while FR I radio galaxies are located in the region below the line of demarcation. This phenomenon implies that the accretion rate may play a crucial role in the classification of FR I and FR II radio galaxies. The reason for the differences in radio morphology and luminosity between these two types of radio galaxies has long been an open question. The cause may be related to the interaction of the jets with the surrounding medium or to the intrinsic properties of the jets and their formation process.

In Chisellini \& Celotti, the mass of the black hole at the centre of the radio galaxy was estimated from, among other things, the absolute optical r-band of the host galaxy, while the jet power was estimated from the radio luminosity at 1.4 GHz. Thus, Ledlow et al. 1996 suggested that the division between FR I and FR II can also be described in the plane of jet power and black hole mass~\cite{ledlow199620cm}. Ghisellini \& Celotti 2001 suggested that the ratio of jet power to black hole mass can be used to classify FR I and FR II radio galaxies~\cite{ghisellini2001dividing}, as they suggested that the difference between FR I and FR II radio galaxies is due to the fact that they are artificial. galaxies have differences rooted in the fact that they have different accretion patterns. In this case, the accretion pattern of FR I radio galaxies belongs to the ADAF branch because the accretion rate of FR I radio galaxies is significantly low~\cite{narayan1995advection(b)}, whereas the accretion pattern of FR II radio galaxies belongs to the standard accretion disc. Under the unified model, FR I radio galaxies are considered to be BL Lacertae (BL Lac) objects with relativistic jets aligned to our line of sight, and FR II radio galaxies correspond to quasars, with BL Lac objects having Eddington ratios systematically lower than those of quasars and in the split of ${L_{bol}}/{L_{Edd}} \sim 0.01$~\cite{xu2009bl}. This suggests that there may be a difference in the two accretion patterns, rather than a difference due to dust ring obscuration.

Cao and Rawlings propose a new explanatory framework for the observed differences between FR I and FR II radio galaxies~\cite{cao2004no}. Based on the Blandford-Znajek jet energy extraction theory (which assumes that the black hole has a very high spin parameter), they calculate the theoretical maximum jet power for a given black hole mass and validate it by combining radio-optical band data from FR I radio galaxies in the 3CR catalogue. The results show that the measured jet power of some of the radio sources exceeds the theoretical prediction by up to an order of magnitude, while the black hole mass distributions between the samples do not show significant differences. Further analyses show that the jet power of FR I radio sources is significantly lower than the observed value if they are in the radial-dominated accretion flows (ADAFs) or adiabatic-dominated outflow (ADIOS) modes. This contradiction suggests that the accretion process of such objects may be dominated by standard thin discs. Furthermore, the linear correlation between the radio luminosity and the intensity of emission lines in the narrow line region can be rationalised by a model in which the central optical core region is obscured by a dust ring. Therefore, the aim of this chapter's study is to investigate the accretion pattern of FR I as well as to verify the presence of a dust ring, and we have chosen 3C 348 as the object of study.

\section{CLOUDY configuration and analysis} \label{sec:2}

The mechanism of emission line formation in active galactic nuclei (AGN) can be traced back to the physical process of photoionisation: high-energy photons emitted by the central engine excite the perinuclear ionised gas cloud, which generates characteristic emission lines through a photoionisation-compound cycle. A central condition for this process is the presence of a source of high-intensity ionising radiation, a role that is assumed in active galactic nuclei by the central supermassive black hole mass accretion system. Based on this principle, we can simulate the radiative transport in the narrow line region by constructing a photoionisation model using the CLOUDY program, which is a common tool for such simulations, to calculate theoretical spectra.

Cao \& Rawlings in 2004 analysed 33 FR I radio galaxies in the 3CR catalogue selected by Chiaberge et al. and revealed a correlation between jet power and accretion patterns, estimating the black hole masses of FR I radio galaxies based on the absolute magnitude of the host galaxy's R-band versus the black hole mass~\cite{mclure2002black}.
\begin{equation}
	{\log _{10}}\left(\frac{{{M_{BH}}}}{{{M_ \odot }}}\right) =  - 0.5(0.02){M_R} - 2.96( \pm 0.48)
\end{equation}

The jet power(${Q_{jet}}$) is estimated from the radio photometry at 151 MHz~\cite{willott1999emission}:
\begin{equation}
	{Q_{jet}} \simeq 3 \times {10^{38}}{f^{3/2}}L_{151}^{6/7}W
\end{equation}

${L_{151}}$ is the total radio luminosity at 151 MHz. The unit is ${10^{28}}W \cdot H{z^{ - 1}} \cdot {s^{ - 1}}$. It is shown that if the accretion mode is assumed to be runoff-dominated accretion modes (ADAFs), then 39\% (13 in total) of the radio sources have measured jet powers that are an order of magnitude higher than the maximum jet power based on theoretical calculations. These 13 FR I radio galaxies are categorised as high jet power (HJP) sources, while the remaining FR I radio galaxies are categorised as low jet power (LJP) sources, and their calculated jet maximum power is shown as a solid line in Fig. There is no difference in the distribution of black hole masses between the HJP sources and the LJP sources, and the presence of the HJP sources therefore suggests that the accretion patterns of FR I radio galaxies may be more complex than previously expected. that was previously expected. This suggests that there is diversity in the accretion patterns of FR I radio galaxies and that HJP sources may be dominated by standard accretion discs rather than ADAF. However, data from Hubble Telescope (HST) observations show that the optical nuclear emission from such sources is weak or even absent, and Cao \& Rawlings suggest that this contradiction can be explained by the dust ring obscuration model - the dust ring obscures the accretion disc emission but preserves the emission in the narrow region - and is supported by the positive correlation between the narrow luminosity and the jet power. relationship also supports this view~\cite{rawlings1991evidence}.

In order to study the properties of HJP radio sources, we have selected one of the high-jet-power FR I radio sources, 3C 348, for this study. The object has a redshift parameter $Z$ = 0.154 and a host galaxy apparent magnitude ${m_v} = 16.9$~\cite{trussoni2001bepposax}. It exhibits peculiar morphological features: although its radio extended structure has the typical bipetal configuration of FR II sources, with a jet extended scale on the order of 500 ks gap and a radio luminosity that exceeds the FR classification value, it lacks the dense hot spot feature inside the radio electronic structure that is the hallmark of FR II sources. At the same time, the object does not fulfil the morphological criteria of a typical FR I source~\cite{dreher1984rings}. Combined with multi-band observations, 3C 348 exhibits a unique low-ionisation state (a very low-ionisation radio galaxy), which contrasts with the high ionisation characteristics common to most high-energy FR II sources~\cite{buttiglione2010optical}.

The [O III] emission line luminosity of the 3C 348 radio source was measured by Buttiglione et al. in 2009, with the result ${L_{[O{\rm{ III}}]}} = {10^{41.29}}erg{~}{{\rm{s}}^{ - 1}}$~\cite{buttiglione2009optical}, and based on the results of Cao and Rawlings in 2004, it was possible to obtain the optical and nuclear luminosities of this radio source, i.e., ${L_{c.opt}} = {10^{30.94}}W/{\AA}$~\cite{willott1999emission}, which translates to a total luminosity of ${10^{35.14}}erg{~}{{\rm{s}}^{ - 1}}$ at an effective wavelength of 6910.33 ${\AA}$. In order to investigate the accretion pattern of the 3C 348 radio source and to determine the presence or absence of dust loops, we carried out photogenic ionisation model calculations using the CLOUDY program~\cite{ferland1998cloudy}. Given that the [O III] emission line is generated by a narrow line, whose flux is not interfered with by the dust loops, the disc luminosity of the optical band of 3C 348, i.e. the endowment optics, is estimated with the help of the observed [O III] line luminosity, and subsequently compared with the actual wavelength of the radio source. We estimate the endowment optical luminosity with the help of the observed [O III] line luminosities, which are subsequently compared with the nuclear optical luminosities obtained from actual observations.

For the CLOUDY model calculations for the 3C 348 radio source, we make the following assumptions about the input parameters:

(1) For the 3C 348 radio source, we consider two representative SED spectral shapes: one is the SED without Big Blue Bump (BBB), which is used to simulate the SED characteristics of the ADAF discs, and the other is the SED with the Big Blue Bump, which simulates the SED characteristics of the standard discs. The SED of the ADAF can be approximated to be characterised as a single power-law continuum with exponential cutoffs at the low frequency end (${10^{12}}H{\rm{z}}$) and at the high frequency end (${10^{20}}Hz$). In the case of the standard disc mode, the template can be represented by a typical ANG continuum with a blue packet temperature of $4.9 \times {10^5}K$, an X-ray to UV ratio of ${\alpha _{ox}} = 1.35$, and a large blue packet continuum with a default low-energy slope of ${\alpha _{uv}} = 0.5$ and a slope of -0.85 for the X-ray component. The broad wavelength thermophotometric range is $1.001 \times {10^{ - 8}}Ryd \sim 7.354 \times {10^6}Ryd$.

(2) The hydrogen density in the narrow line region of the 3C 348 radio source was determined to be according to ${n_H} = {10^4}c{m^{ - 3}}$.

(3) The chemical composition of the gas in the narrow line region we take twice the solar abundance.

(4) The radius of the narrow line region is calculated according to the equation obtained by Liu G et al. (2013) between the scale of the narrow line and the luminosity of the [O III] emission line~\cite{liu2013observations}:
\begin{equation}
	\begin{split}
		{\log _{10}}\left(\frac{{{R_{{\mathop{\rm int}} }}}}{{pc}}\right) = &(0.25 \pm 0.02){\log _{10}}\left(\frac{{{L_{[O~III]}}}}{{{{10}^{42}}erg{~}{{\rm{s}}^{- 1}}}}\right) \\
		& + (3.75 \pm 0.03)
	\end{split}
\end{equation}

Using the CLOUDY photoionisation model, the thermal luminosity was fine-tuned to match the calculated [O III] luminosity to the observations. We obtain a thermal luminosity of ${L_{bol}} = {10^{44.84}}erg{~}{{\rm{s}}^{ - 1}}$ for 3C 348 in both the no-blue-embedded SED and with significant blue-embedded SED models. From the thermal luminosity of 3C 348 we can estimate the monochromatic luminosity corresponding to the wavelength of the centre of the Hubble Telescope (HST) filter in the observer's coordinate system ${L_{cloudy}}$ using the SED related to the accretion rate $ \dot{m}(\dot{m} \equiv \dot{M}/\dot{M}_{Edd})$. Thus the same monochromatic luminosity, ${L_{cloudy}} = {10^{40.31}}erg{~}{{\rm{s}}^{ - 1}}{~}{{\AA}^{ - 1}}$, is obtained for the 3C 348 radio source in the absence of a blue-packet SED and in the presence of a significant blue-packet SED for comparison with its optical core luminosity, ${L_{obs}} = {10^{37.71}}erg{~}{{\rm{s}}^{ - 1}}{~}{{\AA}^{ - 1}}$ (determined using the flux data of Chiaberge et al. (1999)). It can be seen that the observed luminosity of the 3C 348 radio source in the optical band is significantly lower than its core optical luminosity, regardless of the SED used. Therefore, we infer that there is a masking phenomenon in this radio source, and the masking material is most likely a dust ring. The actual observed low nuclear luminosity is due to the fact that the radiation from the central source is obscured by the dust ring. This finding supports the hypothesis proposed by Cao and Rawlings in 2004 that the anomalous observational characteristics of the high-jet-power FR I radio source may be related to the occlusion effect in the line-of-sight direction.

Next we scaled the thermal luminosity to the Eddington luminosity to get the Eddington rate:
\begin{equation}
	\frac{{{L_{bol}}}}{{{L_{Edd}}}} = {10^{ - 3.6}}
\end{equation}

Where ${L_{Edd}} = 1.3 \times {10^{38}}{M_{BH}}/{M_ \odot }erg~{{\rm{s}}^{ - 1}}$, combined with the total 151 MHz radio emission photometric data of the jet to estimate the power of the radio source jets of 3C 348 and others, where the total jet power of 3C 348, ${Q_{jet,t}} = {10^{46.1}}erg{~}{{\rm{s}}^{ - 1}}$, and we also estimate the power of the jet in the nuclear region based on the radio emission of the nuclear region to obtain ${Q_{jet,c}} = {10^{44.2}}erg{~}{{\rm{s}}^{ - 1}}$, which gives ${Q_{jet,t}}/{L_{Edd}} = {10^{ - 1.5}}$, ${Q_{jet,c}}/{L_{Edd}} = {10^{ - 3.4}}$. There is a significant correlation between FR I radio galaxies and FR II radio galaxies in the ${L_{bol}}/{L_{Edd}} - {Q_{jet}}/{L_{Edd}}$ plane, as shown in Fig:
\begin{equation}
	\log ({L_{bol}}/{L_{Edd}}) = (1.03 \pm 0.09)\log ({Q_{jet}}/{L_{Edd}}) - (0.19 \pm 0.27)
\end{equation}

We consider the existence of this correlation for the total jet power lobe of the 3C 348 radio power source, which can be obtained with a centre engine thermometric luminosity of than Eddington luminosity of:
\begin{equation}
	\frac{{{L_{bol,d}}}}{{{L_{Edd}}}} = {10^{ - 1.7}}
\end{equation}

\begin{figure}[!ht]
	\centering
	\includegraphics[width=\linewidth]{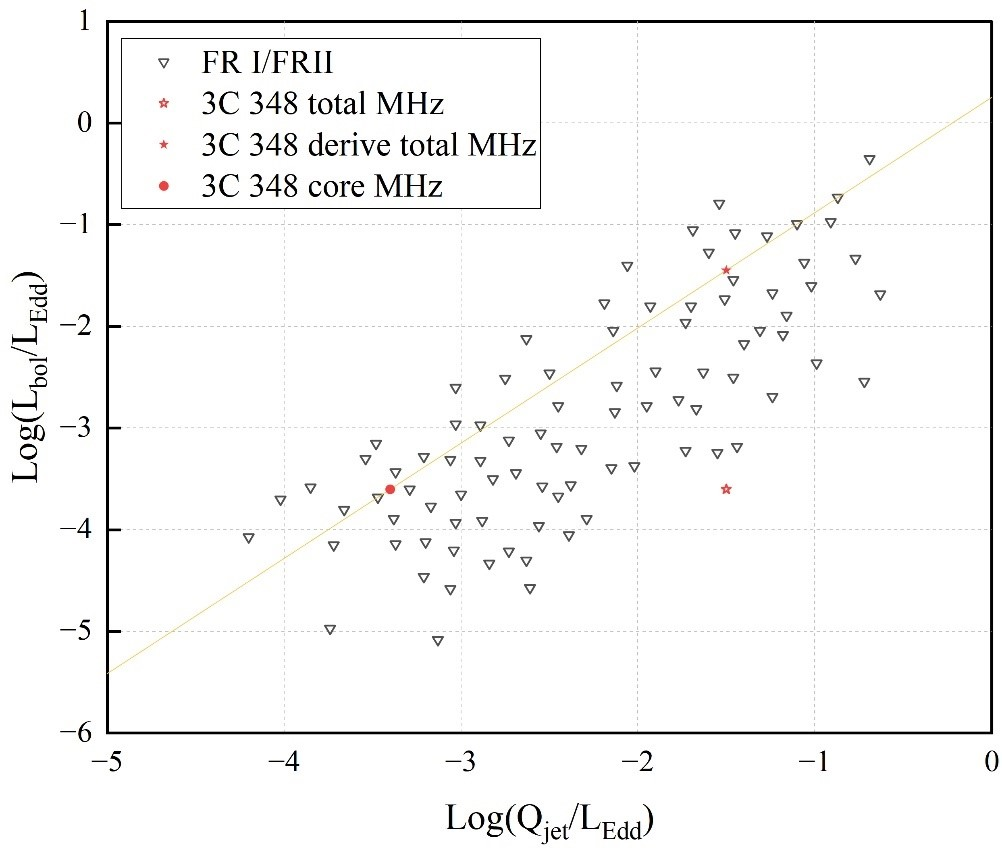}
	\caption{The solid yellow line in the figure indicates the best-fit curve for the other FR I/FR II sources, the red solid circle and hollow pentagram are divided into is the 3C 348 source nuclear region and the total jet power, and the red solid pentagram is the total calorimetric photometry derived from the fit curve.}
	\label{fig:1}
\end{figure}

Based on the 3C 348 thermometric luminosity and the jet power to Eddington luminosity ratio, we can see from the figure that the centre engine of 3C 348 is in a low state, which is consistent with a radial shift-dominated accretion flow (ADAF) signature. We know that the Bondi accretion rate of 3C 348 is $0.089{M_ \odot }y{r^{ - 1}}$ can also be written as $1.6 \times {10^{ - 3}}{\dot{M} _{Edd}}$, where ${\dot{M}_{Edd}} = 2.28{M_{BH}}/{M_8}{M_ \odot }y{r^{ - 1}}$ (${M_8}$ denotes ${10^8}{M_ \odot }$). Based on the characteristics of the ADAF+jet composite model 3C 348 centre engine is investigated, we use the overall solution of ADAF around a rapidly rotating black hole, where the variation of the accretion rate with radius ($R$) can be written as:
\begin{equation}
	\dot{M} = {\dot{M}_{out}}{\left(\frac{R}{{{R_{out}}}}\right)^{{p_w}}}
\end{equation}

\noindent where ${R_{out}}$ is the outer boundary radius, ${\dot{M}_{out}}$ is the accretion rate and pw is the disc wind parameter. We make ${\dot{M}_{out}} = \alpha {\dot{M}_{bondi}}$, where the viscosity coefficient takes the standard value of 0.3, and use the electron turbulence heating ratio $\delta$ (with a standard value of 0.1) and the magnetic field strength ratio $\beta$ (with a standard value of 0.5). In the study of the jet formation mechanism, we make the assumption that for a radial shift-dominated accretion flow (ADAF), only $\dot{M}_{jet}$ fraction of the accreted material is vertically ejected to form a jet structure. Based on the framework of internal excitation theory, we introduce the electron acceleration efficiency ${\xi _e}$ and the power-law spectral index $p$ (the standard values of ${\xi _e}$ and $p$ are 0.05 and 2.5, respectively), and set the energy share of accelerated electrons and magnetic field in the jet to be ${\epsilon_e} = 6\% $ and ${\epsilon_B} = 2\%$ respectively.

For the special properties of 3C 348, the model parameters are set to 0.8 times the speed of light for jet motion, an angle of 50$^\circ$ between the observation line of sight and the jet axial direction, and a standard value of 5$^\circ$ for the jet tensor angle. The interior of the 3C 348 radio source is assumed to have a very high spin parameter, $a$ = 0.99 (${\rm{a}} = J/{M_{BH}}{R_g}c$), of a Kerr black hole with a dimensionless spin determined by the angular momentum and the mass of the black hole. For the analysis of the radiative energy spectrum, the jet power is estimated by a hybrid model driving the jet~\cite{gizani2003multiband,feng2017constraint}:
\begin{equation}
	{Q_{jet}} = B_p^2{R^4}{\Omega ^2}/32c
\end{equation}

\noindent where $R$ denotes the typical scale of the jet-generating region, and ${B_p}$ denotes the polar magnetic field strength, which is regulated by the endowed magnetic field amplification factor g of the accretion disc (${B_p} \simeq g{B_{dynamo}}$). In relativistic accretion flow theory, the physical evolution of the magnetic field amplification factor is dominated by the dynamical coupling of the intrinsic angular momentum ($\Omega$) of the accretion disc to the local spatiotemporal corrected angular momentum ($\Omega^{\prime}$), which is quantitatively characterised as $g = \Omega /{\Omega ^{\prime}}$. The calculation of the modified angular momentum requires the introduction of the metric coupling factor $\omega  =  - {g_{\phi t}}/{g_{\phi \phi }}$, a key parameter for the spatio-temporal drag effect in the Boyer-Lindquist coordinate system. For the model calculations, all core physical quantities are calibrated at the characteristic radius of the innermost stable circular orbit ($R = {R_{ms}}$). Due to the insufficient availability of multi-band observations in the nuclear region (resolution limitations and upper flux constraints in the infrared to X-ray bands), the parameters to be optimised in the theoretical model are reduced to two key degrees of freedom, namely, the disc wind structure index ${p_w}$ and the jet outflow rate ${\dot{M}_{jet}}$.

Fitting of the spectral energy distribution based on the EVN high-resolution radio data shows that the 3C 348 radio source outflow rate ${\dot{M}_{jet}} = {10^{ - 5.2}}{\dot{M}_{Edd}}$, and no significant accretion component is detected. The parameter space analysis shows that the disc wind strength PPP of 3C 348 needs to satisfy greater than 0.1, and when this parameter is increased to 0.4, the ADAF contribution to the energy transport is significantly attenuated, which leads to a simultaneous decrease in the jet power in the nuclear region (kpc scale). Therefore, in the framework of relativistic accretion flow theory, in order to maintain the mass transport balance in the black hole neighbourhood, the numerical simulation strictly limits the disk wind strength parameter ${p_w} \le 0.4$. It is worth noting that under the assumption of zero outflow, the accretion flow needs to reach a critical accretion rate of order ${\dot{M}_{acc}} \ge {10^{-1.5}}{\dot{M}_{Edd}}$ at five times the Swarthy radius scale ($5{R_s}$) to satisfy the energy demand of the large-scale jet power of the 3C 348 radio source.

Taken together, the 3C 348 radio source exhibits a unique accretion feature: the measured Eddington ratio (${L_{bol}}/{L_{Edd}}$) in the nuclear region is consistently below the 1\% order of magnitude, which significantly contradicts the theoretical expectation of over 1\% based on the inversion of the total jet power. It is noteworthy that the critical accretion rate predicted by the theory of phase transitions in the accretion state is in the 1\% Eddington ratio interval. This systematic deviation from observation and theory suggests that the core engine of this radio source may be at a critical evolutionary stage - i.e., the transition from the radiation-dominated standard thin-disk accretion mode to the path-shift-dominated accretion flow (ADAF) evolution!

\section{Spectral decomposition} \label{sec:3}

The analysis of the optical spectra from the Italian National Galileo Telescope (TNG), using optical spectroscopic observations of FR I radio galaxies, enables a study to be carried out with respect to the proportion of the contribution of active galactic nuclei and host galaxies to the observed optical emission.~\cite{buttiglione2010optical} The spectral fitting of the 3C 348 radio source was done by means of the public code PyQSOFit~\cite{guo2018pyqsofit,shen2019sloan}.

\begin{figure}[!ht]
	\centering
	\includegraphics[width=\linewidth]{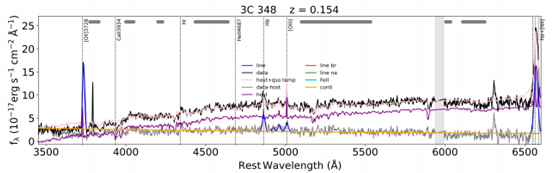}
	\caption{Spectrogram of the 3C 348 radio source. The horizontal coordinates indicate the rest wavelength and the vertical coordinates indicate the spectral flux density. The black curve in the figure is the actual data observed; the pink curve is a temperature-dependent model fit of the host galaxy and quasar (QSO); the purple curve represents the contribution of the host galaxy; the grey curve is the data after subtracting the contribution of the host galaxy The orange curve in the figure represents the continuum spectrum; and the other coloured curves represent the different spectral lines, respectively.}
	\label{fig:2}
\end{figure}

First, a decomposition based on the templates of Yip et al. (2004a) and Yip et al.~\cite{yip2004distributions,yip2004spectral}, (2004b) was performed using a principal component analysis (PCA) method to separate the host galaxy from 3C 348. Next, we fit the pseudo-3C 348 continuum spectrum using a power-law continuum spectrum, except for the emission line regions. The pseudo-continuum model was then subtracted from the 3C 348 component to obtain the emission-line-only spectrum, and this subtracted-continuum spectrum was fitted using a multi-Gaussian model for the emission lines only. The spectrum of the 3C 348 radio source was well decomposed, and in the fitting process we also masked out the atmospheric absorption lines, which are indicated by grey areas in Fig. The results of the fit for 3C 348 are shown in Fig.~\ref{fig:2}, and it is clear that its observed optical emission in the optical wavelength range comes mainly from its host galaxy. This result suggests that the central engine of the 3C 348 radio source is indeed obscured by the dust ring.

\section{Summary} \label{sec:4}

The classification system of radio galaxies has not yet formed a unified standard, and there are various classification bases in the academic community; Fanaro and Riley proposed the classification system of FR I and FR II based on radio morphology and radiation intensity in 1974~\cite{khachikian1974atlas}. In contrast, Liang proposed in 1994 to classify these objects into two categories, low excitation (LEG) and high excitation (HEG), based on optical spectral features~\cite{laing1994spectrophotometry}, with high excitation (HEG) galaxies corresponding to FR II morphology, and low excitation (HEG) galaxies having both FR I and FR II morphology. Although the two types of galaxies do not show significant differences in host galaxy luminosity distributions and black hole masses, it has been shown that the FR morphology classification and spectral classification do not strictly correspond to each other, and that the high and low excitation states cannot be simply equated with FR II and FR I types. This contradiction has prompted the academic community to re-examine the physical nature of radio galaxies, focusing on two main directions: differences in the central engine accretion mechanism (Ghisellini et al. suggest ADAF accretion for FR I versus standard thin-disk accretion for FR II) or the dust obscuration effect (Cao et al. suggest that some of the high jet power of FR I may originate from obscuration of the dust ring in the line-of-sight direction~\cite{cao2004no}). Consequently, Cao and Rawlings in 2004 suggested the presence of dust ring obscuration effects in at least some of the high-jet-power FR I radio sources, since the ADAF accretion model is unable to form jets with such high power in the Blandford-Znajek mechanism. In order to study the physical properties of high jet power FR I radio galaxies, we have chosen the radio source 3C 348 to investigate its central engine accretion model and the presence of dust rings.

Simulations by the CLOUDY photoionisation program reveal that the optical emission from the nuclear region of 3C 348 is significantly lower than the theoretical prediction, suggesting the presence of a dust ring obscuration effect. A later fitting of the 3C 348 spectrogram by spectral decomposition methods shows that its observed optical emission is mainly from the host galaxy, further supporting the obscuration effect. We also found that its nuclear region accretion rate is much lower than the accretion rate inverted from the total jet power through the 3C 348 thermometric photometry and the ratio of the jet power to the Eddington photometry relation, suggesting that its accretion mode may be in the stage of transition from a standard thin disc to a radial shift dominated accretion flow (ADAF). Using its Bondi low accretion rate, its disc wind parameter range of $0.1 \le {p_w} \le 0.4$ is obtained, which further suggests that the shift in the accretion mode may affect the jet power evolution through the energy transfer mechanism.

\bibliography{cite}
\end{document}